# Geometrical Models and Hadronic Radii

Sarwat Zahra[1], Haris Rashid[2], Fazal-e-Aleem[3], Talab Hussain[4], Abrar Ahmad Zafar[4], Sohail Afzal Tahir[4]

[1] University of Education, Township Campus, Lahore - 54770, Pakistan.
[2] Govt. College Women University Sialkot, Pakistan.
[3] The University of Lahore - 54500, Pakistan.
[4] Centre for High Energy Physics, University of the Punjab, Lahore - 54590, Pakistan.

**Abstract:** By using electromagnetic form factors predicted by Generalized Chou Yang model (GCYM), we compute root mean square (rms) radii of several hadrons with varying strangeness content (number of strange quarks/anti-quarks) such as $(\pi, p, \varphi, \Lambda, \Sigma^+, \Sigma^-$ and $\Omega^-)$. The computed radii are found quite consistent with the experimental results and those from other models (for pion and proton). For hadrons other than pion and proton, the experimental results are not available and also the GCYM and other models' results are not consistent with each other. The computed rms radii (from GCYM and other models) indicate that rms radii decrease with increase in strangeness content, separately for mesons and baryons. The experimental results of hadrons other than pion and proton will throw more light on the suitability of GCYM and other models.

**PACS:** 14.20.Gk, 13.40.Gp, 14.40.Df,

## 1. Introduction:

Exploring the elementary particles and their interactions is the essence of particle physics. Study of hadronic matter can provide important information in this regard. Physical picture of the hadronic matter is rather unclear and experimental studies at various experiments, such as TOTEM [1], can throw more light on various aspects in this direction. TOTEM experiment is especially behind exploring the structure of proton which is yet poorly unraveled. In our work, structure of proton and other hadrons has been in focus by computing their rms radii through electromagnetic form factors provided by GCYM. Once, more and more precise TOTEM data come, better comparison can be made with the models such as GCYM. In this regard, the form factors are good tools in explaining elastic and diffractive scattering within the framework of geometrical picture [2-6]. A detailed discussion has been undertaken in Ref. [6] in this regard.

Studying hadronic properties through fundamental constituents; gluons and quarks, is an important challenge for particle physics. Important properties include; total charge, magnetic moment, rms radius, etc. The total charge and magnetic moment are well described by constituent quark framework whereas the charge and current distributions are not well understood. The rms radius of a hadron represents its size, and can be studied through hadronic form factors which are functions of charge and current distributions [7]. The process of quark confinement which is poorly understood, justifies the finite hadronic sizes. The finite rms radii can be measured in elastic electron-hadron scattering [8]. In addition to quark confinement, the quark flavor can also play an important role towards the finite sizes of hadrons. Such as according to experimental measurement [8], the difference in the pion and kaon radii is justified through different strange quark content. This point can be investigated through systematic study of the radii of hyperons with varying strangeness content.

The hadronic sizes can be probed through total, elastic and differential cross sections. Experimental measurements [9] showed that total cross section increases logarithmically with the increasing energy. This was an important discovery that led the geometrical models to employ one of the two possibilities: "(1) size of a hadron increases with increasing energy and (2) the hadronic matter distribution and its size are independent of the energy. In the second case, the increase in total cross section with increasing energy is due to increase in opacity with increase in energy or increase in interaction strength with increase in energy" [10]. In this regard, the matter distribution is assumed to be similar to the charge distribution which is independent of energy. The matter distribution might also affect the cross section at asymptotic energies [10, 11].

Geometrical picture has been a successful tool in describing hadronic processes [2-6]. In this picture, the elastic scattering of hadrons is described with the fundamental assumption that hadrons are extended structures. These extended objects make elastic collisions by passing through each other with some attenuation. The minimal and the extended versions [2-6] of the geometrical model have been very successful in providing consistent formalism for studying several geometrical and physical aspects in hadronic physics. This includes information about the hadronic sizes, hadronic matter distribution, dip structures in the elastic differential cross section, inward movement of dips with increasing energy, possibility of more dips at ultra-high energies, etc. [2, 10, 12].

On the side of geometrical picture of hadrons, Chou and Yang Model [13], was proposed in 1968, to study elastic scattering of hadrons at asymptotic energies. The model was further improved through prediction of first dip in differential cross section, which was seen in the elastic differential cross section of proton-proton scattering at ISR energies [14]. Since then a number of attempts by the different researchers [7] have been made to fit the world data for hadron-hadron elastic scattering. One of the extended versions [2, 15] of geometrical picture is the Generalized Chou Yang Model [16]. It has been successful in explaining several features of elastic scattering, specially the region beyond the diffraction peak [17]. Its generalization is based upon modified form of the opaqueness $\Omega(s,b)$ appearing in the expression of scattering amplitude $T(s, t)$ [16]:

$$T(s,t) = i\int b J_0(b\sqrt{-t})(i - e^{-\Omega(s,b)})db$$

with

$$\Omega(s,b) = K(1-i\alpha)\int \sqrt{-t}\,d\sqrt{-t}\,J_0(b\sqrt{-t})G_X(t)G_Y(t)f(t)/f(0)$$

where K and b are the normalization constant and the impact parameter, respectively. $J_0$ represents the Bessel function having zero order. $G_X(t)$ and $G_Y(t)$ represent the form factors of the colliding hadrons X and Y. $f(t)/f(0)$ denotes the anisotropy function. The extent of anisotropic behavior of partons is represented by this anisotropy function. α is the ratio of real part to imaginary part of the scattering amplitude in the forward direction. By using scattering amplitude, under the normalization conditions, the differential and total cross sections are given as:

$$\frac{d\sigma}{dt} = \pi |T(s,t)|^2, \ \sigma = 4\pi \operatorname{Im} T(s,t=0)$$

The scattering information of hadrons can be used to extract their form factors. Knowledge of the form factors plays a pivotal role in extracting physical picture of hadrons such as their rms radii. The rms radius of a hadron can be obtained from its form factor by using the relation between rms radius and form factor derivative with respect to momentum transfer square when momentum transfer square approaches to zero. The relation is given as:

$$r_{rms}^2 = -6 \frac{dG(q)}{dq}\bigg|_{q=0}$$ [18], where q represents the

momentum transfer square and G is the form factor of a hadron.

Currently, only the nucleon form factors can be measured experimentally because their stable nucleon targets are readily available. A lot of experimental work has been done to study baryons and mesons in the past decades through their electromagnetic form factors [19-24]. In this regard, many theoretical studies have also been made. Using the relativistic constituent-quark model, V.Cauteren et al. [25] investigated electric and magnetic form factors of strange baryons. In 1996, Kims et al. and in 2001, Kubis et al. used the chiral-quark/soliton and the chiral perturbation theory, respectively to study the form factor of $\Lambda$ baryon [26]. In 2010, C. Alexandrou et al [27] reported the electromagnetic form factor of $\Omega^-$ using the Lattice QCD. Yong-Lu Liu and Ming-Qiu Huang, in 2009, investigated the electromagnetic form factors of $\Lambda$ and $\Sigma$ baryons by using the Ioffe-type interpolating currents [28].

## 2. rms Radii of Hadrons

The hadronic radii are important source of information about properties of the constituent quarks [29]. The sizes of hadrons can provide information about the sizes of constituent quarks and confinement forces. The data presented in [29] on the hadronic radii give significant hint about dependence of rms radii on the strangeness content and on the quark masses [29].

Very prominent objective of scattering experiments is to unravel the internal structure of scattering objects. Once the form factor of a hadron is obtained, its structure can be better estimated by studying its important facets such as its radius which tells about the size of hadron itself. The results thus obtained for different energies of the colliding hadrons can be used to check the two possible pictures about their physical behavior as their energy increases. As the Chou Yang model assumes that size of hadrons and their matter distribution are independent of their incoming energy, it can be investigated whether this possibility or the second one or even their mixture is consistent with the experimental data available at different energy values of colliders [30]. In this direction, we have used form factors of some hadrons to compute their radii and have compared the results with the experimental data and with theoretical predictions of other models.

By using the relation between the rms radius of hadrons and their form factors (given above) we computed rms radii of different hadrons based upon the electromagnetic form factors predicted by GCYM (listed in Table 1). The computed values are reported in Table 2 and Table 3, for comparison with the experimental values and those predicted by other models. On the experimental side only rms radii of pion and proton have been measured. In this regard, the comparison with experimental values of hadronic rms radii is very limited. For these two hadrons, the GCYM predicted rms radii and the experimental values are quite consistent. Some of the other attempts made to compute/measure the values of rms radii of different hadrons are highlighted as follows.

In 1981, M. F. Heyn [35] used ρ+ smooth polynomial model to compute radius of pion as 0.836 fm. In 1982, Dally [36] used dipole form fit to the form factor of pion to get its rms radius which is consistent with our predicted value. In 2001, Eschrichi [8] reported the rms charge and strong interaction radii of $\Sigma^-$ based upon $\Sigma^-$-electron scattering data.

| Hadron | Symbol | Form Factor | Reference |
|---|---|---|---|
| Pion | $\pi$ | $G(t) = 1/(1-1.40845\,t)^{4/3}$ | [6] |
| proton | P | $G(t) = 0.645 e^{4t} + 0.33 e^{0.85t} + 0.028 e^{0.22t} + 0.0015 e^{0.05t}$ | [6] |
| Phi | $\phi$ | $G(t) = 0.9 e^{0.75t - 0.55t^2} + 0.1 e^{0.22t}$ | [6] |
| Lambda | $\Lambda$ | $G(t) = 0.64 e^{2.4t} + 0.36 e^{0.37t}$ | [31] |
| Sigma plus | $\Sigma^+$ | $G(t) = 0.8 e^{1.55t} + 0.23 e^{-t^2} - 0.03$ | [31] |
| Sigma minus | $\Sigma^-$ | $G(t) = e^{0.98t + 0.04t^2}$ | [31] |
| Omega minus | $\Omega^-$ | $G(t) = -e^{0.93t + 0.045t^2}$ | [31] |

Table 1: Electromagnetic Form Factors from GCYM

| Hadron | Symbol | Rest Mass (MeV/c²) | Composition | Experimental rms radii (fm) | GCYM predicted rms radii (fm) |
|---|---|---|---|---|---|
| Pion | $\pi^+$ | 134.9766±0.0006 | $u\bar{d}$ | 0.6625 [32] | 0.662373 |
| Proton | P | 938.272 ±0.000023 | uud | 0.862 [33] <br> 0.895±0.018 [34] | 0.815808 |
| Phi | $\Phi$ | 1,019.445±0.020 | $s\bar{s}$ | ---- | 0.403533 |
| Lambda | $\Lambda$ | 1115.683± 0.006 | uds | ---- | 0.624477 |
| Sigma plus | $\Sigma^+$ | 1189.37±0.07 | uus | ---- | 0.538236 |
| Sigma | $\Sigma^-$ | 1197.449±0.030 | dds | ---- | 0.478493 |
| Omega minus | $\Omega^-$ | 1672.45±0.29 | sss | ---- | 0.466126 |

Table 2: Comparison of GCYM and Experimental rms radii of hadrons

In 1990, Povh [37] reported that geometrical picture can interpret high energy hadron-proton interactions and thus can provide information about radii of interacting hadrons.

In most of the cases, the results of other models are in agreement with our computed values using GCYM, as shown in Table 3. In 2001, Eschrichi [8] also predicted the radii of few hadrons including $\Omega^-$ and $\Delta$. Our calculated results agree with their (computed values) for the case of $\Omega^-$.

| Hadron | Symbol | GCYM predicted radii (fm) | rms radii from other models (fm) |
|---|---|---|---|
| Pion | $\pi$ | 0.662373 | 0.65±0.11 [36] |
| Proton | P | 0.815808 | 0.8185 [37] |
| Phi | $\Phi$ | 0.403533 | 0.45 [8] |
| Lambda | $\Lambda$ | 0.624477 | 0.76 [37] |
| Sigma plus | $\Sigma^+$ | 0.538236 | --- |
| Sigma | $\Sigma^-$ | 0.478493 | 0.79 [8] |
| Omega minus | $\Omega^-$ | 0.466126 | ≤ 0.60 [31] |

Table 3: Comparison between GCYM rms radii versus those from other Models

## 3. Conclusions:

Though large number of hadrons is established very well experimentally, size of very few of them has been probed theoretically as well as experimentally. In fact, the structural study of hadrons can be helpful in understanding the structure of atom and its nucleus. The precise structural information of hadrons such as proton is an essential requirement for precision studies of many physical aspects of atoms/nuclei. In this direction the results of TOTEM experiment will throw more light on the structure and size of Proton.

In our work, GCYM provided electromagnetic form factors have been used to predict rms radii values of several hadrons. Because the experimental values of rms radii of hadrons are only available for pion and proton, the comparison with the experimental results is very limited. As far as pion and proton are concerned, the GCYM predicted values of rms radii and those from other models are consistent with each other as well as with the available experimental values. For other hadrons, the GCYM predicted values and those from other models are not consistent with each other. Therefore the experimental measurements of rms radii of other hadrons are very important for concluding whether the GCYM or the other models are better because more is the consistency with the experimental values more better might be a model.

In addition to comparison among GCYM predictions, experimental values and those from other models, an interesting general trend is seen among the GCYM predicted rms radii of hadrons. It can be seen that the rms radii values decrease with increase in strangeness content in the hadrons (separately for mesons and baryons), as shown in Fig. 1 and in Table 2. More interesting thing is that the general trend can also be seen among the rms radii values predicted by other models. Experimental results of rms radii for hadrons other than pion and proton are also very essential for further scrutiny of this important aspect of hadronic radii.

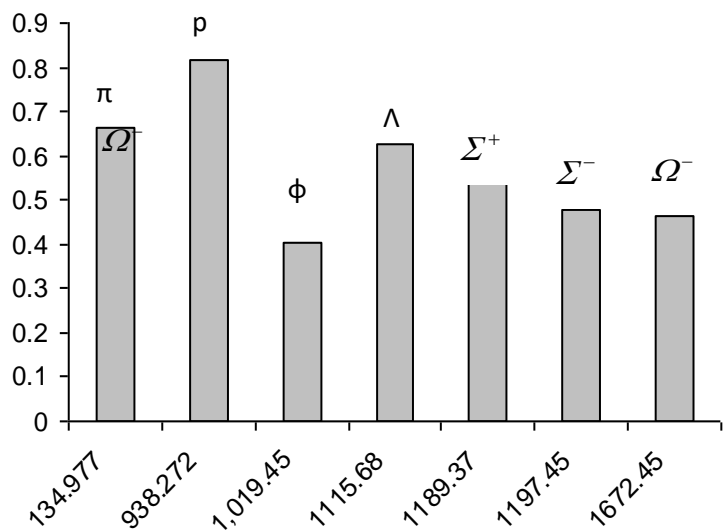

Fig. 1  Masses of Hadrons versus their rms radii